\documentclass[12pt]{article}
\usepackage{amssymb,amsmath,epsfig}
\usepackage{graphicx}
\usepackage[skip=2pt]{caption}
\usepackage{subcaption}
\usepackage{caption}
\usepackage{float}
\usepackage{epstopdf,leftidx,mathtools}
\usepackage{arydshln}

\newcommand{\be}{\begin{equation}}
\newcommand{\ee}{\end{equation}}
\newcommand{\bea}{\begin{eqnarray}}
\newcommand{\eea}{\end{eqnarray}}
\newcommand{\beaa}{\begin{eqnarray*}}
\newcommand{\eeaa}{\end{eqnarray*}}

\newcommand{\BB}{{{\rm I} \kern -2pt \rlap {\rm B} \kern +8pt}}

\DeclareCaptionFormat{myformat}{\fontsize{9}{10}\selectfont#1#2#3}
\advance\textheight by \topskip
\makeatletter

\@addtoreset{equation}{section}

\def\section{\@startsection {section}{1}{\z@}{-3.5ex plus -1ex minus
 -.2ex}{2.3ex plus .2ex}{\large\bf\centering}}
\def\subsection{\@startsection{subsection}{2}{\z@}{-3.25ex plus%
 -1ex minus -.2ex}{1.5ex plus .2ex}{\bf}}
\def\subsubsection{\@startsection{subsubsection}{3}{\z@}{-3.25ex plus%
 -1ex minus -.2ex}{1.5ex plus .2ex}{\sl}}
\makeatother
\begin{document}

\baselineskip 18pt
\parindent 12pt
\parskip 10pt

\begin{center}
{\Large {\bf On soliton solutions of the time-discrete generalized
lattice Heisenberg magnet model }}\\\vspace{1 cm} {\large
H. Wajahat A. Riaz\footnote{%
ahmed.phyy@gmail.com} and Mahmood ul Hassan  \footnote{%
mhassan@physics.pu.edu.pk} }\vspace{0.15in}

{\small{\it Department of Physics, University of Punjab,
Quaid-e-Azam Campus, \\Lahore-54590, Pakistan.}}
\end{center}

\begin{abstract}
Generalized lattice Heisenberg magnet model is an integrable model
exhibiting soliton solutions. The model is physically important for
describing the magnon bound states (or soliton excitations) with
arbitrary spin, in magnetic materials. In this paper, a
time-discrete generalized lattice Heisenberg magnet (GLHM) model is
investigated. By writing down the Lax pair representation of the
time-discrete GLHM model, we present explicitly the underlying
integrable structure like, the Darboux transformation and soliton
solutions.
\end{abstract}

\vspace{0.5 cm} \textit{PACS: 02.30.Ik, 05.45.Yv
\\   Keywords: Integrable systems,
solitons, Discrete Darboux transformation}

\section{Introduction}
Discrete or (lattice) integrable systems namely systems with their
independent variables defined on a lattice points have received much
attention by the researchers working in the field of theoretical and
applied sciences. The study of discrete integrable system not only
as a physical model but also in the context of numerical analysis is
of particular importance in various fields ranging from pure
mathematics to experimental science. Discrete integrable systems
such as, Toda lattice, Volterra lattice, Ablowitz-Ladik lattice,
Hirota-Miwa equation, nonlinear $\sigma$-model, sine-Gordon equation
etc have been studied extensively in the literature
\cite{suris}-\cite{skylin}. Soliton solutions of these discrete
systems have been computed by using various systematic methods such
inverse scattering transform, B{\"a}cklund/Darboux transformation,
Hirota bilinear method etc \cite{suris}-\cite{skylin}, \cite{21}.

The lattice Heisenberg magnet model has been studied in many
references such as \cite{faddeev}-\cite{haldane}. It exhibits many
aspects of integrability for instance Lax pair representation,
higher symmetries, $r$-matrix formulism etc. The soliton solutions
have been studied by B{\"a}cklund transformation (BT), Darboux
transformation (DT), inverse scattering transform (IST) and other
solution generating techniques \cite{faddeev}-\cite{enej}.

The Lax pair representation of the time-discrete GLHM model is given
by \cite{tsuchida}
\begin{eqnarray}
\Phi_{n+1}^{m}&=&\mathcal A_{n}^{m}\Phi_{n}^{m}, \qquad \mathcal A_{n}^{m}=I+\lambda \mathcal U^{m}_{n}, \label{lax1} \\
\Phi^{m+1}_{n}&=&\mathcal B^{m}_{n}\Phi^{m}_{n}, \qquad \mathcal
B^{m}_{n}=I + h \frac{\lambda}{1-\lambda^{2}}\mathcal J^{m}_{n} + h
\frac{\lambda^{2}}{1-\lambda^{2}}\mathcal J^{m}_{n}\mathcal
U^{m}_{n}, \label{lax2}
\end{eqnarray}
where $\mathcal U^{m}_{n}$ is an $N \times N$ matrix and
$\Phi_{n}^{m}$ is also an $N \times N$ eigen-function matrix. The
conditions on the matrix $\mathcal U^{m}_{n}$ i.e., $ \left(\mathcal
U^{m}_{n}\right)^{2} = I$ and $\mathcal J^{m}_{n} \mathcal A^{m}_{n}
= \mathcal A^{m}_{n-1} \mathcal J^{m}_{n}$ are assumed. The latter
condition implies $\mathcal J^{m}_{n} \mathcal U^{m}_{n}= \mathcal
U^{m}_{n-1}\mathcal J^{m}_{n}$. The compatibility condition of the
Lax pair (\ref{lax1})-(\ref{lax2}) implies a zero-curvature
condition i.e., $\mathcal A^{m+1}_{n}\mathcal B^{m}_{n} = \mathcal
B^{m}_{n+1} \mathcal A^{m}_{n}$ which is equivalent to the equation
of motion
\begin{equation}\label{E}
\frac{1}{h}\left[\left(\mathcal A_{n}^{m+1}\right)^{-1} -
\left(\mathcal A_{n}^{m}\right)^{-1}\right] +
\frac{\lambda}{1-\lambda^{2}}\left(\mathcal J^{m}_{n+1} - \mathcal
J^{m}_{n} \right) = O,
\end{equation}
or equivalently,
\begin{equation}\label{EOM}
\frac{1}{h}\left(\mathcal U_{n}^{m+1} - \mathcal U_{n}^{m}\right) =
\mathcal J^{m}_{n+1} - \mathcal J^{m}_{n}
\end{equation}
The relation $\mathcal J^{m}_{n} \mathcal U^{m}_{n} = \mathcal
U^{m+1}_{n-1}\mathcal J^{m}_{n}$ is satisfied if we choose $\mathcal
J^{m}_{n} = 2 \dot{\imath} a_{n}^{m} \mathcal
U^{m+1}_{n-1}\left(\mathcal U^{m}_{n} + \mathcal
U^{m+1}_{n-1}\right)^{-1} + 2b_{n}^{m} \left(\mathcal U^{m}_{n} +
\mathcal U^{m+1}_{n-1}\right)^{-1}$. Substituting this expression
into equation (\ref{EOM}), we obtain
\begin{equation}\label{EOM1}
\frac{1}{h}\left(\mathcal U_{n}^{m+1} - \mathcal U_{n}^{m}\right) =
\Delta^{+}_{n}\left[2 \dot{\imath} a^{m}_{n} \mathcal
U^{m+1}_{n-1}\left(\mathcal U^{m}_{n} + \mathcal
U^{m+1}_{n-1}\right)^{-1} + 2b_{n}^{m} \left(\mathcal U^{m}_{n} +
\mathcal U^{m+1}_{n-1}\right)^{-1}\right],
\end{equation}
where $\Delta_{n}^{+} f^{m}_n = f^{m}_{n+1} - f^{m}_{n}$. For $N=2$,
we get a simplest $2 \times 2$ case and express the equation
(\ref{EOM1}) as
\begin{equation}\label{EOMSU2}
\frac{1}{h}\left(\textbf{U}_{n}^{m+1} - \textbf{U}_{n}^{m}\right) =
\Delta^{+}_{n}\left[a \frac{\textbf{U}^{m}_{n} \times
\textbf{U}^{m+1}_{n-1}}{1 +
\textbf{U}^{m}_{n}.\textbf{U}^{m+1}_{n-1}} + b_{n}^{m}
\frac{\textbf{U}_{n}^{m} + \textbf{U}^{m+1}_{n-1}}{1 +
\textbf{U}^{m}_{n}.\textbf{U}^{m+1}_{n-1}}\right],\qquad
\left(\textbf{U}^{m}_{n}\right)^{2} = 1.
\end{equation}
It should be noted that the generalization of the lattice Heisenberg
model (\ref{EOM}), (\ref{EOM1}) was studied by Tsuchida
\cite{tsuchida}. In the case of matrices of size $2 \times 2$, it
reduces to the well-known vector lattice Heisenberg chain
(\ref{EOMSU2}). The B{\"a}cklund transformations for the latter are
sufficiently well studied, in particular, their derivation is given
in \cite{tsuchida}, with references to earlier works. However, the
problem of describing B{\"a}cklund/Darboux transformations for the
general matrix case (\ref{EOM}), (\ref{EOM1}) is left open
\cite{tsuchida}. It is this problem that is considered in the
present work.

In this paper, we present a systematic approach to find the soliton
solutions of the time-discrete GLHM model (\ref{EOM}). We define
Darboux transformation (DT) on the solution to the Lax pair and the
solutions of the matrix generalization of the time-discrete GLHM
model defined by equation (\ref{EOM}) with respect to the $N \times
N$ matrices $\mathcal J_{n}^{m},\; \mathcal U_{n}^{m}$ , or, after
reduction, by one equation (\ref{EOM1}) with respect to the matrix
$\mathcal U_{n}^{m}$.

Darboux transformation is one of the powerful and effective
technique used to compute solutions of a given nonlinear integrable
equation in soliton theory. The main idea of this method is that, a
new solution to the Lax pair (i.e., pair of linear equations
associated with nonlinear integrable equation) can be obtain from
the old solution by means of Darboux matrix. The covariance of the
Lax pair under the Darboux transformation requires that the new
solution satisfies the same Lax pair such that the relationship
between new and old solutions to the Lax pair and the solutions to
the nonlinear integrable equation can be built. Hence, one can find
the soliton solutions to the nonlinear integrable equation by
solving a Lax pair with the given seed (or trivial) solutions.
Various integrable equations have been studied successfully by means
of Darboux transformation and obtained the soliton solutions have
been computed. The solutions are expressed in terms of Wronskian,
quasi-Grammian and quasi-determinants in the literature
\cite{5}-\cite{6}, \cite{21}-\cite{shi}.

This paper is organized as follows. Section 2, contains the
derivation of the DT for the matrix generalization of the
time-discrete GLHM model (\ref{EOM}), (\ref{EOM1}). Furthermore, the
solutions obtained by DT are expressed in qusideterminant form. In
section 3, soliton solutions for the general $N \times N$ and
simplest $N=2$ case of the time-discrete GLHM model are obtained.
Section 4, is devoted for concluding remarks.

\section{Discrete Darboux transformation}\label{section2}
In what follows, we apply DT on the Lax pair
(\ref{lax1})-(\ref{lax2}) of the time-discrete GLHM model
(\ref{EOM}) to obtain soliton solutions. We define a DT on the
solutions to the Lax pair equations (\ref{lax1})-(\ref{lax2}) by
means of a $N \times N$ discrete Darboux matrix $D_{n}^{m}$. The
discrete Darboux matrix $D_{n}^{m}$ acts on the solution
$\Phi_{n}^{m}$ of the Lax pair (\ref{lax1})-(\ref{lax2}) to give
another solution $\Phi_{n}^{m}[1]$ i.e.,
\begin{equation}\label{si1}
\Phi_{n}^{m}[1]=D^{m}_{n}\Phi^{m}_{n}.
\end{equation}
The covariance of the Lax pair (\ref{lax1})-(\ref{lax2}) under the
DT requires that the new solution $\Phi_{n}^{m}[1]$ satisfies the
same Lax pair equations with the new matrices $\mathcal
A_{n}^{m}[1],\; \mathcal B_{n}^{m}[1]$ i.e.,
\begin{eqnarray}
\mathcal A^{m}_{n}[1] &=& I + \lambda \mathcal U^{m}_{n}[1],
\label{darbouxlax1}
\\
\mathcal B^{m}_{n}[1]&=&I + h\frac{\lambda}{1-\lambda^{2}}\mathcal
J^{m}_{n}[1] + h \frac{\lambda^{2}}{1-\lambda^{2}}\mathcal
J^{m}_{n}[1]\mathcal U^{m}_{n}[1], \label{darbouxlax2}
\end{eqnarray}
By using (\ref{lax1})-(\ref{lax2}), equations
(\ref{darbouxlax1})-(\ref{darbouxlax2}) imply that the discrete
Darboux matrix $D_{n}^{m}$ satisfies the following discrete
Darboux-Lax equations as
\begin{equation}\label{darbouxmatrixequation}
\mathcal A_{n}^{m}[1] D_{n}^{m} = D_{n+1}^{m} \mathcal A_{n}^{m},
\qquad \mathcal B_{n}^{m}[1] D_{n}^{m} = D_{n}^{m+1} \mathcal
B_{n}^{m}
\end{equation}
We are interested in finding a DT on the matrices $\mathcal
U_{n}^{m}[1],\;\mathcal J^{m}_{n}[1]$. For this, we make the ansatz
for the Darboux matrix such as $D_{n}^{m}=\lambda^{-1}{I}-\mathcal
Q^{m}_{n},$ where $Q^{m}_{n}$ is the $N \times N$ auxiliary matrix
and $I$ is the $N \times N$ identity matrix. By substituting the
latter expression of $D_{n}^{m}$ in equation
(\ref{darbouxmatrixequation}), the coefficients of $\lambda$ yields
the DT on the matrices $\mathcal U^{m}_{n}$, $\mathcal J^{m}_{n}$ as
\begin{eqnarray}
\mathcal U^{m}_{n}[1] &=& \mathcal U^{m}_{n} - (\mathcal Q^{m}_{n+1} - \mathcal Q^{m}_{n}), \label{u[1]} \\
\mathcal J^{m}_{n}[1] &=& \mathcal J^{m}_{n} -
\frac{1}{h}\left(\mathcal Q^{m+1}_{n} - \mathcal Q^{m}_{n} \right),
\label{j[1]}
\end{eqnarray}
with the following conditions on the Darboux matrix $\mathcal
Q^{m}_{n}$
\begin{eqnarray}\label{a10}
\left(\mathcal Q^{m}_{n+1} - \mathcal Q^{m}_{n}\right) \mathcal Q^{m}_{n} &=& \mathcal U^{m}_{n} \mathcal Q^{m}_{n}
- \mathcal Q^{m}_{n+1} \mathcal U^{m}_{n}, \label{condition1} \\
\frac{1}{h}\left(\mathcal Q_{n}^{m+1} - \mathcal Q_{n}^{m}
\right)\left(I- (\mathcal Q^{m}_{n})^{2}\right) &=&\left[\mathcal
Q^{m}_{n},\; \mathcal J_{n}^{m} \left(\mathcal Q^{m}_{n} + \mathcal
U^{m}_{n}\right) \right]^{+}. \label{condition2}
\end{eqnarray}
where $\left[f_{n}^{m},\;g_{n}^{m}\right]^{+} :=
f_{n}^{m+1}g_{n}^{m} - g_{n}^{m}f_{n}^{m}$. In what follows, we show
that for $\mathcal Q_{n}^{m} = \Theta_{n}^{m} \Lambda^{-1}
(\Theta_{n}^{m})^{-1}$, where $\Theta_{n}^{m}$ is the $N \times N$
particular matrix solution to the Lax pair (\ref{lax1})-(\ref{lax2})
which is constructed by $N$ wave function $\Phi_{n}^{m}$ for
different values of $\lambda$, whereas the matrix $\Lambda$ is an $N
\times N$ diagonal matrix with $N$ distinct eigenvalues
$\lambda_{i}$ ($\lambda_{i} \neq \lambda_{j}$),
$i=1,\;...,\;N,\;j=i+1,\;...,\;i+N$. Take $N$ constant column basis
vectors $\left|e_{1}\right\rangle,\;...,\;
\left|e_{N}\right\rangle$, so that the invertible $N \times N$
matrix $\Theta_{n}^{m}$ can be defined as $\Theta_{n}^{m} =
\left(\Phi_{n}^{m}(\lambda_{1}) \left|e_{1}\right\rangle,\;...,\;
\Phi_{n}^{m}(\lambda_{N})\left|e_{N}\right\rangle\right) =
\left(\left|\theta_{1}\right\rangle_{n}^{m},\;...,\;
\left|\theta_{N}\right\rangle_{n}^{m}\right)$, such that each
$\left|\theta_{i}\right\rangle_{n}^{m}=
\Phi_{n}^{m}(\lambda_{i})\left|e_{i}\right\rangle$ in the matrix
$\Theta_{n}^{m}$ is a column solution to the Lax pair
(\ref{lax1})-(\ref{lax2}). For
$\lambda=\lambda_{i}\;(i=1,\;...,\;N)$, we have
\begin{eqnarray}
\left|\theta_{i}\right\rangle^{m}_{n+1} &=&
\left|\theta_{i}\right\rangle^{m}_{n} +
\lambda_{i} {\mathcal U}^{m}_{n}\left|\theta_{i}\right\rangle^{m}_{n}, \label{columnlax1} \\
\left|\theta_{i}\right\rangle^{m+1}_{n} &=&
\left|\theta_{i}\right\rangle^{m}_{n} + h
\frac{\lambda_{i}}{1-\lambda_{i}^{2}}{\mathcal J}^{m}_{n}
\left|\theta_{i}\right\rangle^{m}_{n} + h \frac{\lambda_{i}^{2}}
{1-\lambda_{i}^{2}}\mathcal J_{n}^{m} \mathcal
U^{m}_{n}\left|\theta_{i}\right\rangle^{m}_{n}, \label{columnlax2}
\end{eqnarray}
For $\Lambda= \text{diag}(\lambda_{1},\;...,\;\lambda_{N})$, the Lax
pair (\ref{columnlax1})-(\ref{columnlax2}) reduce to the generalized
matrix form as
\begin{eqnarray}
\Theta^{m}_{n+1} &=& \Theta^{m}_{n} + \mathcal U^{m}_{n} \Theta^{m}_{n} \Lambda, \label{Glax} \\
\Theta^{m+1}_{n} &=& \Theta^{m}_{n} + h \mathcal J^{m}_{n}
\Theta^{m}_{n}\Lambda \left(I-\Lambda^{2}\right)^{-1} + h \mathcal
J^{m}_{n} \mathcal U^{m}_{n} \Theta^{m}_{n}
\Lambda^{2}\left(I-\Lambda^{2}\right)^{-1}, \label{Glax2}
\end{eqnarray}
where $\Theta^{m}_{n}$ is a particular matrix solution to the Lax
pair (\ref{lax1})-(\ref{lax2}). Let us define a matrix $\mathcal
Q^{m}_{n}$ in terms of an invertible matrix $\Theta^{m}_{n}$, i.e.
$\mathcal Q^{m}_{n} = \Theta^{m}_{n} \Lambda^{-1}
(\Theta^{m}_{n})^{-1}$. Now, we show that the latter expression of
the matrix $\mathcal Q_{n}^{m}$ satisfies the set of equations
(\ref{condition1})-(\ref{condition2}). To do this, let us check the
first condition as
\begin{eqnarray}\label{a16}
&&\left(\mathcal Q^{m}_{n+1} - \mathcal Q^{m}_{n} \right) \mathcal Q^{m}_{n}, \notag \\
&&= \left(\Theta^{m}_{n+1}\Lambda^{-1}
\left(\Theta^{m}_{n+1}\right)^{-1}
- \Theta^{m}_{n} \Lambda^{-1} \left(\Theta^{m}_{n}\right)^{-1}\right)\Theta^{m}_{n} \Lambda^{-1} \left(\Theta^{m}_{n}\right)^{-1}, \notag \\
&&= \Theta^{m}_{n+1} \Lambda^{-1}
\left(\Theta^{m}_{n+1}\right)^{-1}\Theta^{m}_{n} \Lambda^{-1}
\left(\Theta^{m}_{n}\right)^{-1} - \Theta^{m}_{n} \Lambda^{-2}
\left(\Theta^{m}_{n}\right)^{-1} + \Theta^{m}_{n+1} \Lambda^{-2}
\left(\Theta^{m}_{n}\right)^{-1}
- \Theta^{m}_{n+1} \Lambda^{-2} \left(\Theta^{m}_{n}\right)^{-1}, \notag \\
&&= \left(\Theta^{m}_{n+1} - \Theta^{m}_{n}\right) \Lambda^{-1}
\left(\Theta^{m}_{n}\right)^{-1} \Theta^{m}_{n} \Lambda^{-1}
\left(\Theta^{m}_{n}\right)^{-1} - \Theta^{m}_{n+1} \Lambda^{-1}
\left(\Theta^{m}_{n+1}\right)^{-1}\left(\Theta^{m}_{n+1}
- \Theta^{m}_{n}\right) \Lambda^{-1} \left(\Theta^{m}_{n}\right)^{-1}, \notag \\
&&= \mathcal U^{m}_{n} \mathcal Q^{m}_{n} - \mathcal Q^{m}_{n+1}
\mathcal U^{m}_{n},
\end{eqnarray}
which is equation (\ref{condition1}). Similarly for the second
condition, we have
\begin{eqnarray}\label{a17}
&&\frac{1}{h}\left(\mathcal Q^{m+1}_{n} - \mathcal
Q_{n}^{m}\right)\left(I -
\left(\mathcal Q^{m}_{n}\right) ^{2}\right) \notag \\
&& = \frac{1}{h} \left(\Theta^{m+1}_{n} \Lambda^{-1}
\left(\Theta^{m+1}_{n}\right)^{-1} - \Theta^{m}_{n} \Lambda^{-1}
\left(\Theta^{m}_{n}\right)^{-1} + \Theta^{m+1}_{n} \Lambda^{-1}
\left(\Theta^{m}_{n}\right)^{-1} - \Theta^{m+1}_{n} \Lambda^{-1}
\left(\Theta^{m}_{n}\right)^{-1}\right) \notag \\
 && \quad \times \Theta_{n}^{m} \left(I-\Lambda^{-2}\right)
\left(\Theta_{n}^{m}\right)^{-1}, \notag
\\
&& = \frac{1}{h} \left(\Theta^{m+1}_{n} \Lambda^{-1}
\left(\Theta^{m+1}_{n}\right)^{-1}\left(\Theta_{n}^{m} -
\Theta_{n}^{m+1}\right) - \left(\Theta^{m}_{n} -
\Theta_{n}^{m+1}\right) \Lambda^{-1} \right)
\left(I-\Lambda^{-2}\right) \left(\Theta_{n}^{m}\right)^{-1}, \notag
\\
&&= -\Theta^{m+1}_{n} \Lambda^{-1}
\left(\Theta^{m+1}_{n}\right)^{-1}\left({\mathcal
J}^{m}_{n}\Theta^{m}_{n} \Lambda \left(I-\Lambda^{2}\right)^{-1} +
\mathcal J^{m}_{n} \mathcal U^{m}_{n}
\Theta^{m}_{n}\Lambda^{2} \left(I-\Lambda^{2}\right)^{-1}\right) \left(I-\Lambda^{-2}\right) \left(\Theta^{m}_{n}\right)^{-1} \notag \\
&& \quad + \left({\mathcal J}^{m}_{n}\Theta^{m}_{n} \Lambda
\left(I-\Lambda^{2}\right)^{-1} + \mathcal J^{m}_{n} \mathcal
U^{m}_{n} \Theta^{m}_{n}\Lambda^{2}
\left(I-\Lambda^{2}\right)^{-1}\right) \Lambda^{-1}
\left(I-\Lambda^{-2}\right) \left(\Theta^{m}_{n}\right)^{-1}, \notag \\
&& = \left[\mathcal Q^{m}_{n},\; \mathcal J^{m}_{n}\left(\mathcal
Q^{m}_{n} + \mathcal S^{m}_{n}\right)\right]^{+}.
\end{eqnarray}
which is equation (\ref{condition2}). Therefore, we have established
that the DT on the matrix solutions $\Phi_{n}^{m},\;\mathcal
U_{n}^{m}$ is given by
\begin{eqnarray}
\Phi^{m}_{n}[1] &=& \left(\lambda^{-1} I - \Theta^{m}_{n} \Lambda^{-1} \left(\Theta^{m}_{n}\right)^{-1}\right)\Phi^{m}_{n},  \label{darboux1} \\
\mathcal U^{m}_{n}[1] &=& \mathcal U^{m}_n -\left(\Theta^{m}_{n+1}
\Lambda^{-1} \left(\Theta^{m}_{n+1}\right)^{-1} -
\Theta^{m}_{n}\Lambda^{-1} \left(\Theta^{m}_{n}\right)^{-1}\right), \notag \\
&=& \Theta^{m}_{n+1}\Lambda^{-1} \left(\Theta^{m}_{n+1}\right)^{-1}
\mathcal U^{m}_{n} \Theta^{m}_{n} \Lambda^{-1}
\left(\Theta^{m}_{n}\right)^{-1}.  \label{darboux2}
\end{eqnarray}
At this stage, we can say that DT (\ref{darboux1})-(\ref{darboux2})
preserves the system i.e., if $\Phi_{n}^{m},\;\mathcal U_{n}^{m}$
are the solutions to the Lax pair (\ref{lax1})-(\ref{lax2}) and GLHM
model (\ref{EOM}) respectively, then $\Phi_{n}^{m}[K],\;\mathcal
U_{n}^{m}[K]$ (that correspond to multi-soliton solutions) are also
the solutions of the same equations. The DT (\ref{darboux2}) is also
consistent with the reduction (\ref{EOM1}). For $\mathcal Q_{n}^{m}
= \Theta_{n}^{m} \Lambda^{-1} \left(\Theta_{n}^{m}\right)^{-1}$, it
seems appropriate here to express the solutions
$\Phi_{n}^{m}[K],\;\mathcal U_{n}^{m}[K]$ in terms of
quasideterminants.\footnote{In this paper, we will use
quasideterminants that are expanded about $n \times n$ matrix. The
quasideterminant expression of $N \times N$ expanded about $n \times
n$ matrix is given as
\begin{equation}
 \left \vert \begin{array}{cc}
M_{11} & M_{12} \\
M_{21} & \fbox{$M_{22} $}%
\end{array}%
\right\vert=M_{22} - M_{21}M_{11}^{-1}M_{12}.
\end{equation}
For further details see \cite{riaz2}-\cite{20}.} The matrix solution
$\Phi_{n}^{m}[1]$ to the Lax pair (\ref{lax1})-(\ref{lax2}) with the
particular matrix solution $\Theta_{n}^{m}$ in terms of
quasideterminant can be expressed as
\begin{eqnarray}\label{quasi1}
\Psi_{n}^{m}[1] &\equiv& D_{n}^{m}\Psi^{m}_{n} = \left(\lambda^{-1}
I - \Theta^{m}_{n}\Lambda^{-1}
\left(\Theta^{m}_{n}\right)^{-1}\right)\Phi^{m}_{n},
\notag \\
&=&\left\vert
\begin{array}{cc}
\Theta^{m}_{n} & \Phi^{m}_{n} \\
\Theta^{m}_{n} \Lambda^{-1} & {\fbox{$\lambda^{-1} \Phi^{m}_{n} $}}
\end{array}%
\right\vert .
\end{eqnarray}
And the one-fold Darboux transformation on the matrix field
$\mathcal U^{m}_{n}$ of the time-discrete GLHM is
\begin{eqnarray}\label{quais2}
\mathcal U^{m}_{n}[1] &=& \mathcal Q^{m}_{n+1} \mathcal U^{m}_{n}
\left(\mathcal Q^{m}_{n}\right)^{-1} = \Theta^{m}_{n+1} \Lambda^{-1}
\left(\Theta^{m}_{n+1}\right)^{-1} \mathcal U^{m}_{n}
\left(\Theta^{m}_{n}\Lambda^{-1}
\left(\Theta^{m}_{n}\right)^{-1}\right)^{-1},
\notag \\
&=&\left\vert
\begin{array}{cc}
\Theta^{m}_{n+1} & I \\
\Theta^{m}_{n+1} \Lambda^{-1} & \fbox{$ O $}
\end{array}%
\right\vert \mathcal U^{m}_{n} \left\vert
\begin{array}{cc}
\Theta^{m}_{n} & I \\
\Theta^{m}_{n} \Lambda^{-1} & {\fbox{$O $}}
\end{array}%
\right\vert^{-1}.
\end{eqnarray}
where $O$ is the $N \times N$ null matrix and $I$ is the $N \times
N$ identity matrix. The results obtained in
(\ref{quasi1})-(\ref{quais2}) can be extended and generalized to
$K$-fold DT. For the matrix solutions $\Theta_{k}$ at
$\Lambda=\Lambda_{k}$ $\left(k=1,\;2,\;...,\;K\right)$ to the Lax
pair (\ref{lax1})-(\ref{lax2}), the $K$-times repeated DT $
\Phi_{n}^{m}[K]$ in terms of quasideterminant is written as
\begin{eqnarray}\label{quasi3}
\Phi^{m}_{n}[K] &=& \prod^{K}_{k=1}\left(\lambda{I}-\mathcal
Q^{m}_{n}[K-k]\right)\Phi^{m}_{n}, \notag \\
&=&\prod^{K}_{k=1}\left(\lambda{I}-\Theta^{m}_{n}[K-k] \Lambda_{K-k}^{-1} \left(\Theta^{m}_{n}[K-k]\right)^{-1}\right)\Phi^{m}_{n}, \notag \\
&=&\left\vert
\begin{array}{ccccc}
\Theta^{m}_{n,\;1} & \Theta^{m}_{n,\;2} & \cdots & \Theta^{m}_{n,\;K} & \Phi_{n}^{m} \\
\Theta^{m}_{n,\;1} \Lambda_{1}^{-1} & \Theta^{m}_{n,\;2}
\Lambda_{2}^{-1} & \cdots &
\Theta^{m}_{n,\;K} \Lambda_{K}^{-1} & \lambda^{-1}\Phi_{n}^{m} \\
\vdots & \vdots & \ddots & \vdots & \vdots \\
\Theta^{m}_{n,\;1} \Lambda_{1}^{-K+1} & \Theta^{m}_{n,\;2}
\Lambda_{2}^{-K+1} & \cdots &
\Theta^{m}_{n,\;K} \Lambda_{K}^{-K+1} & \lambda^{-K+1}\Phi_{n}^{m} \\
\Theta^{m}_{n,\;1}\Lambda_{1}^{-K} & \Theta^{m}_{n,\;2}
\Lambda_{2}^{-K} & \cdots & \Theta^{m}_{n,\;K} \Lambda_{K}^{-K} &
\fbox{$\lambda^{-K}\Phi_{n}^{m}$}
\end{array}%
\right\vert.
\end{eqnarray}
Similarly the quasideterminant expression for $\mathcal
U^{m}_{n}[K]$ is
\begin{eqnarray}\label{quasi4}
\mathcal U^{m}_{n}[K] &=& \left\vert
\begin{array}{ccccc}
\Theta^{m}_{n+1,\;1} & \Theta^{m}_{n+1,\;2} & \cdots & \Theta^{m}_{n+1,\;K} & I \\
\Theta^{m}_{n+1,\;1} \Lambda_{1}^{-1} & \Theta^{m}_{n+1,\;2}
\Lambda_{2}^{-1} & \cdots &
\Theta^{m}_{n+1,\;K} \Lambda_{K}^{-1} & O \\
\vdots & \vdots & \ddots & \vdots & \vdots \\
\Theta^{m}_{n+1,\;1} \Lambda_{1}^{-K+1} & \Theta^{m}_{n+1,\;2}
\Lambda_{2}^{-K+1} & \cdots &
\Theta^{m}_{n+1,\;K} \Lambda_{K}^{-K+1} & O \\
\Theta^{m}_{n+1,\;1}\Lambda_{1}^{-K} & \Theta^{m}_{n+1,\;2}
\Lambda_{2}^{-K} & \cdots & \Theta^{m}_{n+1,\;K} \Lambda_{K}^{-K} &
\fbox{$O$}
\end{array}%
\right\vert \notag \\
&\times & \mathcal U^{m}_{n} \left\vert
\begin{array}{ccccc}
\Theta^{m}_{n,\;1} & \Theta^{m}_{n,\;2} & \cdots & \Theta^{m}_{n,\;K} & I \\
\Theta^{m}_{n,\;1} \Lambda_{1}^{-1} & \Theta^{m}_{n,\;2}
\Lambda_{2}^{-1} & \cdots &
\Theta^{m}_{n,\;K} \Lambda_{K}^{-1} & O \\
\vdots & \vdots & \ddots & \vdots & \vdots \\
\Theta^{m}_{n,\;1} \Lambda_{1}^{-K+1} & \Theta^{m}_{n,\;2}
\Lambda_{2}^{-K+1} & \cdots &
\Theta^{m}_{n,\;K} \Lambda_{K}^{-K+1} & O \\
\Theta^{m}_{n,\;1}\Lambda_{1}^{-K} & \Theta^{m}_{n,\;2}
\Lambda_{2}^{-K} & \cdots & \Theta^{m}_{n,\;K} \Lambda_{K}^{-K} &
\fbox{$O$}
\end{array}%
\right\vert^{-1}.
\end{eqnarray}

Equations (\ref{quasi3}) and (\ref{quasi4}) represent respectively,
the required $K$th quasideterminant solutions $\Phi^{m}_{n}[K]$ to
the Lax pair and $\mathcal U^{m}_{n}[K]$ of the time-discrete GLHM
model. These results can be easily proved by induction.

\section{Soliton solutions}\label{section4}

In this section, we obtain the soliton solutions from a seed
(trivial) solution by solving the Lax pair of the time-discrete GLHM
model. For this, we re-write the matrix $\left(\mathcal
Q^{m}_{n}\right)^{(K)}$ from (\ref{quasi4}) in a more convenient
form as follows
\begin{equation}\label{b1}
{\mathcal Q^{m}_{n}}^{(K)} = \left\vert
\begin{array}{cc}
\mathcal G^{m}_{n} & \mathcal{I}^{(K)} \\
\widetilde{\mathcal G}^{m}_{n}&\fbox{$O$}
\end{array}%
\right\vert,
\end{equation}
where $\mathcal{I}^{(K)},\; \widetilde{\mathcal G}^{m}_{n}$ and
$\mathcal G^{m}_{n}$ are $NK \times N,\;N \times NK$ and $NK \times
NK$ matrices respectively. These matrices are given by
\begin{eqnarray}
\mathcal{I}^{(K)} &=& \left(\begin{array}{cccc}
              I & O & \cdots & O
            \end{array}\right)^{T},\notag \\
\widetilde{\mathcal G}^{m}_{n} &=& \left(\begin{array}{cccc}
\Theta^{m}_{n,\;1}\Lambda^{-K}_{1} &
\Theta^{m}_{n,\;2}\Lambda^{-K}_{2}&\cdots&
\Theta^{m}_{n,\;K}\Lambda^{-K}_{K}
\end{array}\right),\nonumber \\
\mathcal G^{m}_{n}&=&\left(
  \begin{array}{cccc}
    \Theta^{m}_{n,\;1} & \Theta^{m}_{n,\;2} & \cdots & \Theta^{m}_{n,\;K} \\
    \Theta^{m}_{n,\;1} \Lambda_{1}^{-1} & \Theta^{m}_{n,\;2} \Lambda_{2}^{-1} & \cdots & \Theta^{m}_{n,\;K} \Lambda_{K}^{-1} \\
    \vdots & \vdots & \ddots & \vdots \\
    \Theta^{m}_{n,\;1}\Lambda^{-K+1}_{1} & \Theta^{m}_{n,\;2}\Lambda^{-K+1}_{2} & \cdots & \Theta^{m}_{n,\;K}\Lambda^{-K+1}_{K} \\
  \end{array}
\right).
\end{eqnarray}
And the components of the matrix $ {\mathcal Q^{m}_{n}}^{(K)}$ can
be decomposed as
\begin{eqnarray}\label{b2}
{\mathcal Q^{m}_{n,\;ij}}^{(K)} &=& \left(\left\vert
\begin{array}{cc}
\mathcal G^{m}_{n} & \mathcal I^{(K)} \\
\widetilde{\mathcal G}^{m}_{n} & \fbox{$O$}
\end{array}%
\right\vert\right)_{ij}=\left\vert
\begin{array}{cc}
\mathcal G^{m}_{n} & \mathcal I^{(K)}_{j} \\
\left(\widetilde{\mathcal
G}^{m}_{n}\right)_{i}&\fbox{$0$}\end{array}
\right\vert, \nonumber \\
&=& - \frac{\det{(\mathcal G^{m}_{n})_{ij}}}{\det{(\mathcal
G^{m}_{n})}},\;\;i,\;j=1,\;2,\;...,\;K.
\end{eqnarray}
where $(\widetilde{\mathcal G}^{m}_{n})_i,\;\mathcal{I}^{(K)}_{j}$
represent $i$-th row and $j$-th column of the matrices
$\widetilde{\mathcal G}^{m}_{n},\;\mathcal{I}^{(K)}$ respectively.
For the simplest matrix of size $2 \times 2$, the matrix ${\mathcal
Q^{m}_{n}}^{(K)}$ can be expressed as
\begin{eqnarray}\label{b5}
{\mathcal Q^{m}_{n}}^{(K)} \equiv \left(
     \begin{array}{cc}
       {\mathcal Q^{m}_{n,\;11}}^{(K)} & {\mathcal Q^{m}_{n,\;12}}^{(K)} \\
       {\mathcal Q^{m}_{n,\;21}}^{(K)} & {\mathcal Q^{m}_{n,\;22}}^{(K)} \\
     \end{array}
   \right) = \left\vert
\begin{array}{cc}
\mathcal G^{m}_{n} & \mathcal{I}^{(K)}\\
\widetilde{\mathcal G}^{m}_{n} & \fbox{$O_{2}$}
\end{array}%
\right\vert,
\end{eqnarray}
with the elements given by
\begin{eqnarray}\label{b6}
{\mathcal Q^{m}_{n,\;ij}}^{(K)} &=& \left\vert
\begin{array}{cc}
\mathcal G^{m}_{n} & \mathcal{I}^{(K)}_{j}\\
(\widetilde{\mathcal G}^{m}_{n})_{i} & \fbox{$0$}
\end{array}
\right\vert = - \frac{\det{(\mathcal
G^{m}_{n})_{ij}}}{\det{(\mathcal G^{m}_{n})}},\;\;i,\;j=1,\;2.
\end{eqnarray}
For one soliton $K=1$, we have
\begin{eqnarray}\label{b7}
\mathcal{I}^{(1)} &=& I_{2} = \left(
                    \begin{array}{cc}
                      1 & 0 \\
                      0 & 1 \\
                    \end{array}
                  \right), \quad
\mathcal G^{m}_{n} = \Theta_{n,\;1}^{m} = \left(
              \begin{array}{cc}
                {\theta^{m}_{n,\;11}}^{(1)} & {\theta^{m}_{n,\;12}}^{(2)} \\
                {\theta^{m}_{n,\;21}}^{(1)} & {\theta^{m}_{n,\;22}}^{(2)} \\
              \end{array}
            \right), \quad
\Lambda_{1}=\left(
                \begin{array}{cc}
                  \lambda_{1} & 0 \\
                  0 & \bar{\lambda}_{1} \\
                \end{array}
              \right), \notag \\
\widetilde{\mathcal G}^{m}_{n} &=& \Theta^{m}_{n,\;1}
\Lambda_{1}^{-1} = \left(
     \begin{array}{cc}
       \lambda_{1}^{-1} {\theta^{m}_{n,\;11}}^{(1)} & \bar{\lambda}_{1}^{-1} {\theta^{m}_{n,\;12}}^{(2)} \\
       \lambda_{1}^{-1} {\theta^{m}_{n,\;21}}^{(1)} & \bar{\lambda}_{1}^{-1} {\theta^{m}_{n,\;22}}^{(2)} \\
     \end{array}
   \right).
\end{eqnarray}
By using equation (\ref{b7}) in (\ref{b6}), the matrix element
${\mathcal Q^{M}_{n,\;12}}^{(1)}$ of the matrix $\mathcal Q^{m}_{n}$
can be computed as
\begin{eqnarray}\label{b8}
{\mathcal Q^{m}_{n,\;12}}^{(1)} &=& \left\vert
\begin{array}{cc}
\mathcal G^{m}_{n} & \mathcal{I}^{(1)}_{2} \\
(\widetilde{\mathcal G}^{m}_{n})_{1} & \fbox{$O_2$}
\end{array}%
\right\vert = \left\vert
\begin{array}{ccc}
{\theta^{m}_{n,\;11}}^{(1)} & {\theta^{m}_{n,\;12}}^{(2)} & 0 \\
{\theta^{m}_{n,\;21}}^{(1)} & {\theta^{m}_{n,\;22}}^{(2)} & 1 \\
\lambda_{1}^{-1} {\theta^{m}_{n,\;11}}^{(1)} &
\bar{\lambda}_{1}^{-1} {\theta^{m}_{n,\;12}}^{(2)} & \fbox{0}
\end{array}%
\right\vert,  \\
&=& - \frac{\det\left(
              \begin{array}{cc}
              {\theta^{m}_{n,\;11}}^{(1)} & {\theta^{m}_{n,\;12}}^{(2)} \\
            \lambda_{1}^{-1} {\theta^{m}_{n,\;11}}^{(1)} & \bar{\lambda}_{1}^{-1}{\theta^{m}_{n,\;12}}^{(2)} \\
                                    \end{array}
                                  \right)}{\det\left(
                                                 \begin{array}{cc}
                                                   {\theta^{m}_{n,\;11}}^{(1)} & {\theta^{m}_{n,\;12}}^{(2)}
                                                   \\
{\theta^{m}_{n,\;21}}^{(1)} & {\theta^{m}_{n,\;22}}^{(2)}
                                                 \end{array}
                                      \right)} = \frac{\left(\lambda_{1}^{-1} - \bar{\lambda}_{1}^{-1}
\right){\theta^{m}_{n,\;11}}^{(1)}{\theta^{m}_{n,\;12}}^{(2)}}
{{\theta^{m}_{n,\;11}}^{(1)}{\theta^{m}_{n,\;22}}^{(2)} -
{\theta^{m}_{n,\;12}}^{(2)}{\theta^{m}_{n,\;21}}^{(1)}}. \notag
\end{eqnarray}
Likewise,
\begin{eqnarray}\label{b9}
{\mathcal Q^{m}_{n,\;21}}^{(1)} = -\frac{\left(\lambda_{1}^{-1} -
\bar{\lambda}_{1}^{-1}\right){\theta^{m}_{n,\;21}}^{(1)}{\theta^{m}_{n,\;22}}^{(2)}}
{{\theta^{m}_{n,\;11}}^{(1)} {\theta^{m}_{n,\;22}}^{(2)} -
{\theta^{m}_{n,\;12}}^{(2)} {\theta^{m}_{n,\;21}}^{(1)}}.
\end{eqnarray}
Similarly, we have
\begin{eqnarray}\label{c4}
{\mathcal Q^{m}_{n,\;11}}^{(1)} &=& - \frac{\lambda^{-1}_{1}
{\theta^{m}_{n,\;11}}^{(1)} {\theta^{m}_{n,\;22}}^{(2)} -
\bar{\lambda}^{-1}_{1}{\theta^{m}_{n,\;12}}^{(2)}
{\theta^{m}_{n,\;21}}^{(1)}} {{\theta^{m}_{n,\;11}}^{(1)}
{\theta^{m}_{n,\;22}}^{(2)} -
{\theta^{m}_{n,\;12}}^{(2)} {\theta^{m}_{n,\;21}}^{(1)}}, \notag \\
\mathcal Q^{(1)}_{n,\;22} &=& - \frac{\bar{\lambda}^{-1}_{1}
{\theta^{m}_{n,\;11}}^{(1)} {\theta^{m}_{n,\;22}}^{(2)} -
{\lambda}^{-1}_{1}{\theta^{m}_{n,\;12}}^{(2)}
{\theta^{m}_{n,\;21}}^{(1)}} {{\theta^{m}_{n,\;11}}^{(1)}
{\theta^{m}_{n,\;22}}^{(2)} - {\theta^{m}_{n,\;12}}^{(2)}
{\theta^{m}_{n,\;21}}^{(1)}}.
\end{eqnarray}
To obtain an explicit form of the soliton solution for the general
$N \times N$ case, let us take $U_{0}\equiv U_{n}=
          \dot{\imath} \left(\begin{smallmatrix}
             c_1 &  \\
            & \ddots &\\
             &  & c_N \\
          \end{smallmatrix}\right), \; a^{m}_{n} = 0, \; b^{m}_{n} =1$ as a seed solution, where $c_{i}$ are real constants and $\text{Tr}(U_{n}) = 0$, so that
solution to the Lax pair (\ref{lax1})-(\ref{lax2}) is given by
\begin{equation}\label{N1}
\Phi^{m}_{n}=\left(
        \begin{array}{cc}
          {\mathcal Z^{m}_{n}}_{p} & O \\
           O & {\mathcal Z^{m}_{n}}_{N-p} \\
        \end{array}
      \right),
\end{equation}
where ${\mathcal Z^{m}_{n}}_{p} = \text{diag}\left(
            {\zeta}_1(\lambda) ,\;...,\; \zeta_{p}(\lambda)
        \right), \; {\mathcal Z^{m}_{n}}_{N-p} = \text{diag}\left(
            \zeta_{p+1}(\lambda),\;...,\; \zeta_{N}(\lambda)
        \right)$
are $p \times p$ and $(N-p) \times (N-p)$ matrices respectively,
whereas $_{n}$ and $^{m}$ appearing in the latter expressions,
denote the discrete indices. And
\begin{equation}
\zeta_{i}(\lambda) = \left(1+\dot{\imath}c_{i}\lambda\right)^{n}
\left(1-\dot{\imath}hc_{i}^{-1}\frac{\lambda}{1-\lambda^{2}} +
h\frac{\lambda^{2}}{1-\lambda^{2}}\right)^{m}.
\end{equation}
For the matrix of size $2 \times 2$, take $U_{0}\equiv \mathcal
U^{m}_{n}=\dot{\imath}\left(
          \begin{smallmatrix}
            c &  \\
             &  -c \\
          \end{smallmatrix}
        \right), \; a_{n}^{m} = 0, \; b^{m}_{n} =1$ as the seed solution, so that a trivial calculation yields
a matrix solution $\Phi_{n}^{m}$ of the Lax pair
(\ref{lax1})-(\ref{lax2}), given by
\begin{equation}\label{a28}
\Phi^{m}_{n}=\left(
        \begin{array}{cc}
          \zeta(\lambda) &  \\
           & \bar{\zeta}(\lambda) \\
        \end{array}
      \right),
\quad \zeta(\lambda) = \left(1+\dot{\imath}c\lambda\right)^{n}
\left(1-\dot{\imath}hc^{-1}\frac{\lambda}{1-\lambda^{2}} +
h\frac{\lambda^{2}}{1-\lambda^{2}}\right)^{m}.
\end{equation}
From equation (\ref{a28}), the $2 \times 2$ matrix $\Theta_{n}^{m}$,
as a particular solution to the Lax pair (\ref{lax1})-(\ref{lax2})
can be constructed as follows
\begin{eqnarray}\label{a30}
\Theta^{m}_{n} =
\left(\Phi^{m}_{n}(\lambda_{1})\left|1\right\rangle,\;\Phi^{m}_{n}(\bar{\lambda}_{1})\left|2\right\rangle\right)
= \left(
     \begin{array}{cc}
       \zeta (\lambda_{1}) & -\zeta (\bar{\lambda}_{1}) \\
       \bar{\zeta} (\lambda_{1}) & \bar{\zeta} (\bar{\lambda}_{1}) \\
     \end{array}
   \right).
\end{eqnarray}
On substituting the matrix $\Theta^{m}_{n}$ in equation (\ref{b5})
with (\ref{b8})-(\ref{c4}), we obtain the expression of the matrix
${\mathcal Q^{m}_{n}}^{(1)}$, given by
\begin{equation}\label{a31}
{\mathcal Q^{m}_{n}}^{(1)}=-\frac{1}{ {\mathcal X_{n}^{m}}^{(+)} +
{\mathcal X_{n}^{m}}^{(-)}}\left(
                                           \begin{array}{cc}
                                             \lambda_{1}^{-1}{\mathcal X_{n}^{m}}^{(+)} + \bar{\lambda}_{1}^{-1}{\mathcal X_{n}^{m}}^{(-)}
                                              & \left(\lambda_{1}^{-1} - \bar{\lambda}_{1}^{-1}\right) {\mathcal Y_{n}^{m}}^{(+)} \\
                                             \left(\lambda_{1}^{-1} - \bar{\lambda}_{1}^{-1}\right) {\mathcal Y_{n}^{m}}^{(-)}
                                             & \bar{\lambda}_{1}^{-1}{\mathcal X_{n}^{m}}^{(+)} + {\lambda}_{1}^{-1}{\mathcal X_{n}^{m}}^{(-)} \\
                                           \end{array}
                                         \right),
\end{equation}
where
\begin{eqnarray}\label{a32}
&& {\mathcal X_{n}^{m}}^{(\pm)} = \left(1 \pm
\dot{\imath}c\lambda_{1}\right)^{n} \left(1 \mp
\dot{\imath}c\bar{\lambda}_{1}\right)^{n} \left(1 \mp
\dot{\imath}h\frac{c^{-1} \lambda_{1}}{1-\lambda_{1}^{2}} +
h\frac{\lambda^{2}_{1}}{1-\lambda^{2}_{1}}\right)^{m} \left(1 \pm
\dot{\imath}h\frac{c^{-1}\bar{\lambda}_{1}}{1-\bar{\lambda}_{1}^{2}}
+ h\frac{\bar{\lambda}_{1}^{2}}{1-\bar{\lambda}_{1}^{2}}\right)^{m},
\notag \\
&& {\mathcal Y_{n}^{m}}^{(\pm)} = \left(1 \pm
\dot{\imath}c\lambda_{1}\right)^{n} \left(1 \pm
\dot{\imath}c\bar{\lambda}_{1}\right)^{n} \left(1 \mp
\dot{\imath}h\frac{c^{-1} \lambda_{1}}{1-\lambda_{1}^{2}} +
h\frac{\lambda_{1}^{2}}{1-\lambda_{1}^{2}}\right)^{m} \left(1 \mp
\dot{\imath}h\frac{c^{-1}\bar{\lambda}_{1}}{1-\bar{\lambda}_{1}^{2}}
+ h\frac{\bar{\lambda}^{2}_{1}}{1-\bar{\lambda}_{1}^{2}}\right)^{m}.
\notag
\end{eqnarray}
From (\ref{quais2}) and (\ref{a31}), we present a one-soliton
solution given by
\begin{equation}\label{a34}
{\mathcal U_{n}^{m}}[1]=\left(
             \begin{array}{cc}
             {u_{n}^{m}} & {v_{n}^{m}}^{(+)} \\
               {v_{n}^{m}}^{(-)} & - {u_{n}^{m}} \\
             \end{array}
           \right),
\end{equation}
where
\begin{eqnarray}\label{a35}
&& {u_{n}^{m}} = \dot{\imath}c + \left(\lambda_{1}^{-1} -
\bar{\lambda}_{1}^{-1}\right)\frac{{\mathcal
X_{n+1}^{m}}^{(+)}{\mathcal X_{n}^{m}}^{(-)} - {\mathcal
X_{n+1}^{m}}^{(-)}{\mathcal X_{n}^{m}}^{(+)}}{\left({\mathcal
X_{n+1}^{m}}^{(+)} + {\mathcal
X_{n+1}^{m}}^{(-)}\right)\left({\mathcal X_{n}^{m}}^{(+)} +
{\mathcal X_{n}^{m}}^{(-)}\right)} , \label{usoliton} \\
&&{v_{n}^{m}}^{\pm} = \left(\lambda_{1}^{-1} -
\bar{\lambda}_{1}^{-1}\right)\frac{{\mathcal
Y^{m}_{n+1}}^{(\pm)}\left({\mathcal X_{n}^{m}}^{(+)} + {\mathcal
X_{n}^{m}}^{(-)}\right) - {\mathcal
Y^{m}_{n}}^{(\pm)}\left({\mathcal X_{n+1}^{m}}^{(+)} + {\mathcal
X_{n+1}^{m}}^{(-)}\right)} {\left({\mathcal X_{n+1}^{m}}^{(+)} +
{\mathcal X_{n+1}^{m}}^{(-)}\right)\left({\mathcal X_{n}^{m}}^{(+)}
+ {\mathcal X_{n}^{m}}^{(-)}\right)}. \notag \label{vsoliton} \\
\end{eqnarray}
\begin{figure}[H]
\centering
        \begin{subfigure}{0.45\textwidth}
                \includegraphics[width=\linewidth]{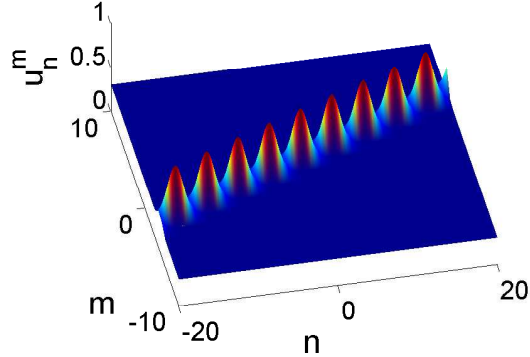}
                \caption{}
        \end{subfigure}
        \caption{Propagation of discrete one-soliton solution (\ref{usoliton}) with the choice of parameters: $c=-0.3,\;\lambda_{1}=1.2\dot\imath$.}
        \label{fig:figure1}
\end{figure}

\begin{figure}[H]
        \centering
        \begin{subfigure}[l]{0.48\textwidth}
                \includegraphics[width=\linewidth]{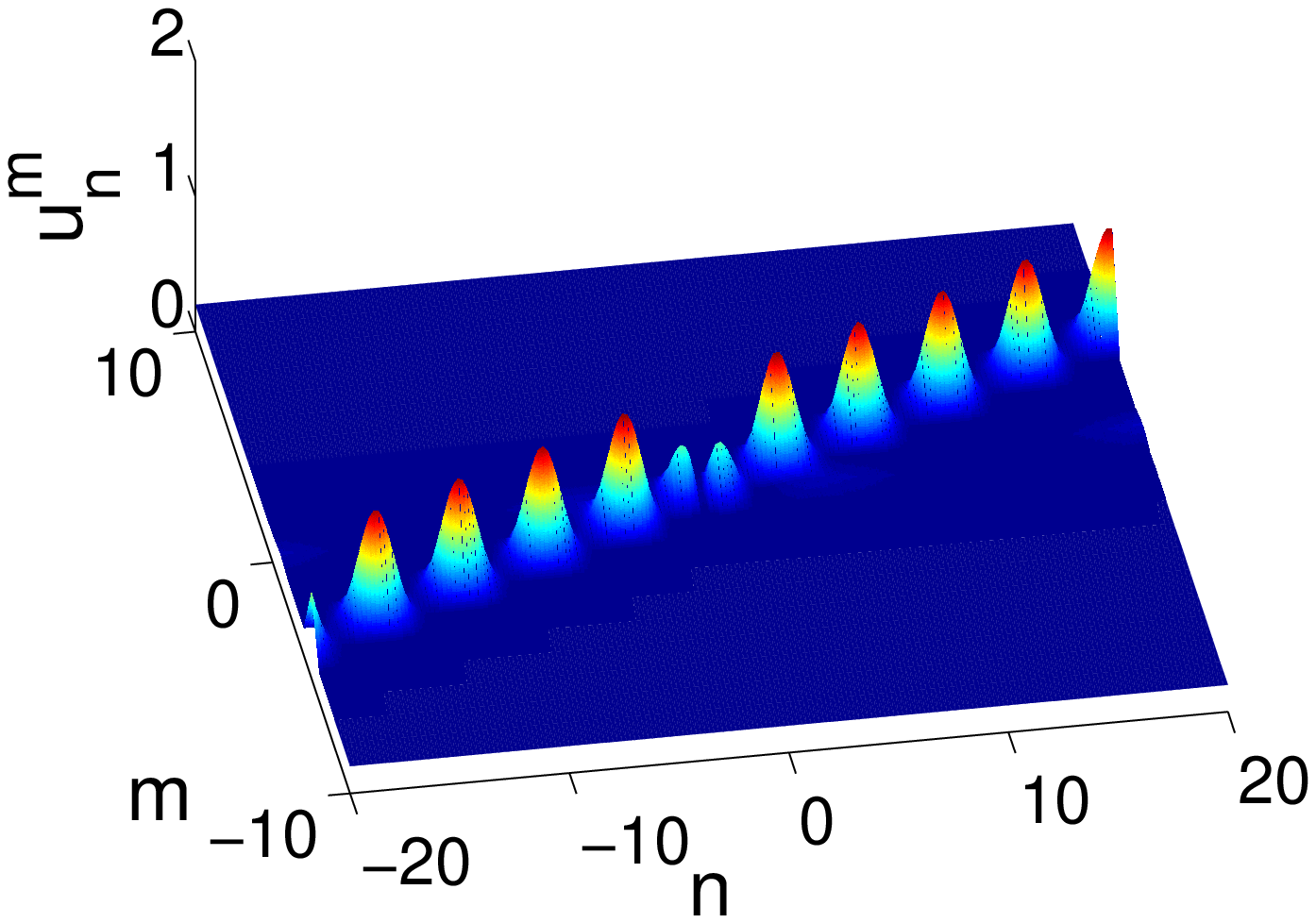}
                \caption{}
        \end{subfigure}
        \begin{subfigure}[r]{0.48\textwidth}
                \includegraphics[width=\linewidth]{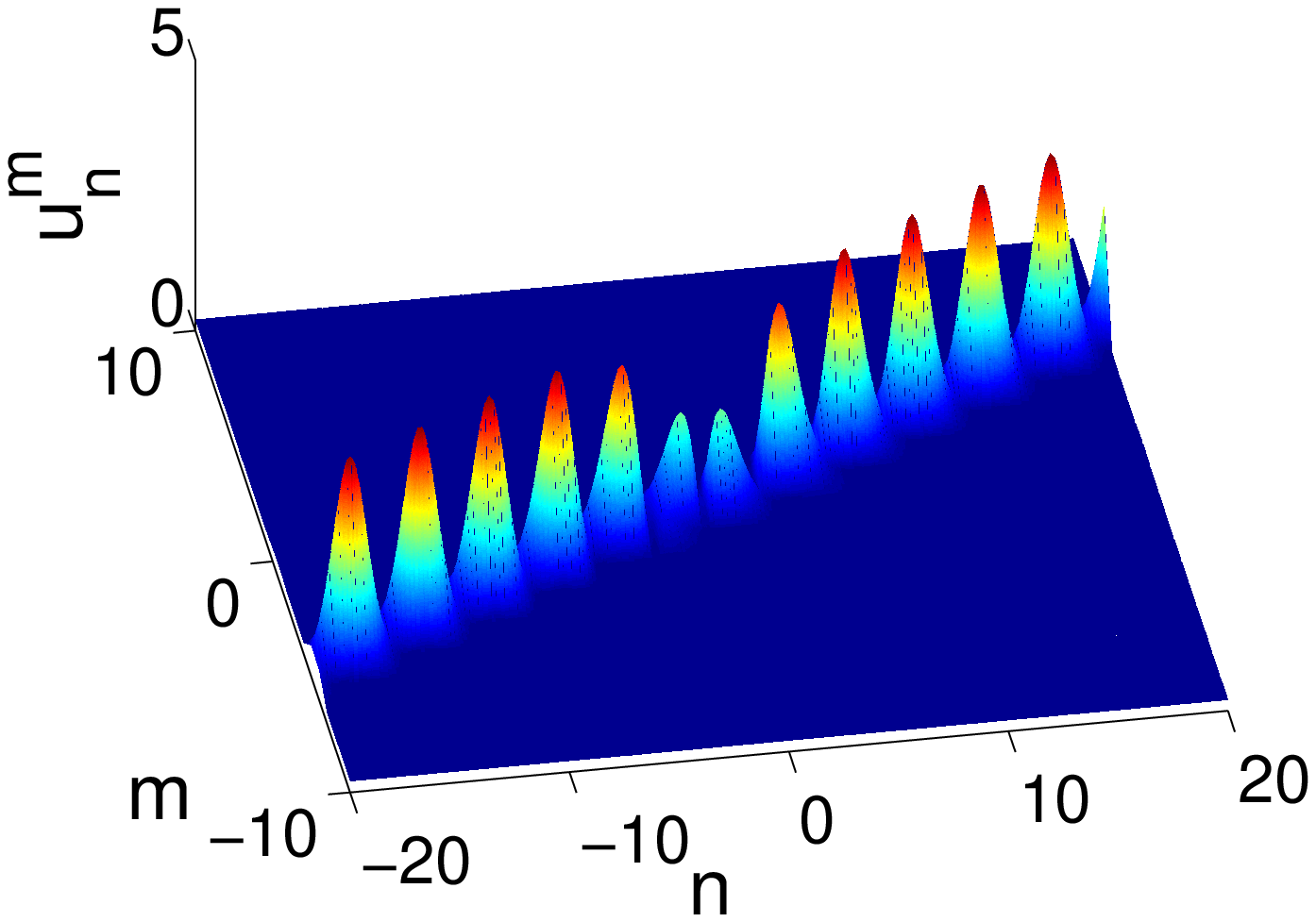}
                \caption{}
        \end{subfigure}
        \caption{Two-soliton solution (\ref{twofoldsol1}). For figure (2a) parameters are; $\lambda_{1}=0.4\dot\imath,\; \lambda_{2}=2.15 \dot\imath
         ,\;c=-0.2$ and for figure (2b): $\lambda_{1}=2.3\dot\imath, \; \lambda_{2} =
         0.1\dot\imath,\;c=-0.2$.}
        \label{fig:figure2}
\end{figure}

\begin{figure}[H]
        \centering
        \begin{subfigure}[l]{0.48\textwidth}
                \includegraphics[width=\linewidth]{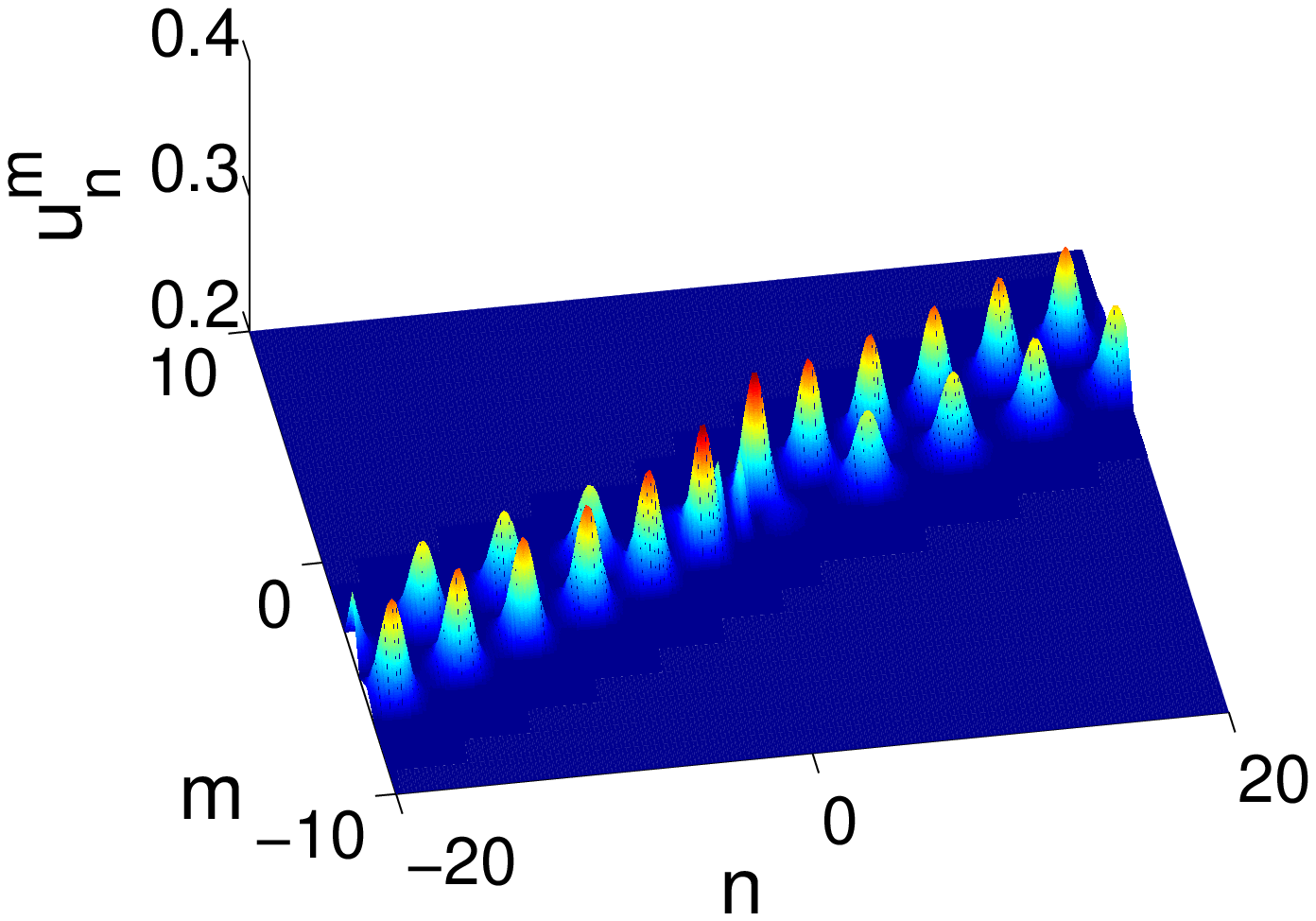}
                \caption{}
        \end{subfigure}
        \begin{subfigure}[r]{0.48\textwidth}
                \includegraphics[width=\linewidth]{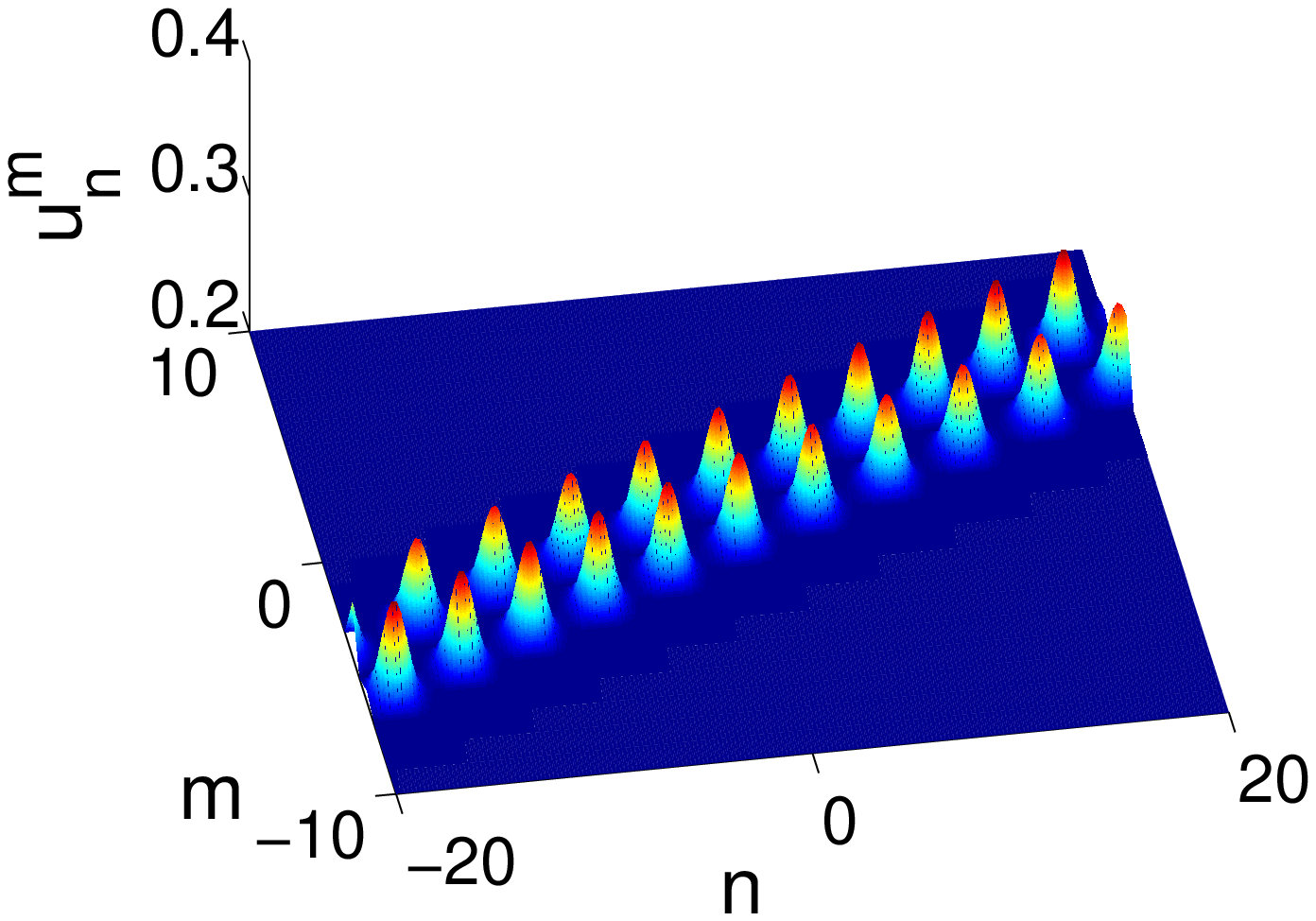}
                \caption{}
        \end{subfigure}
        \caption{Two-soliton solution (\ref{twofoldsol1}). Figure (3a) describe the scattering of two discrete solitons with the choice
        of parameters: $\lambda_{1}=2.34\dot\imath,\; \lambda_{2}=2.15 \dot\imath
         ,\;c=-0.2$ and figure (3b) describe the propagation of two discrete parallel solitons with parameters: $\lambda_{1}=-2.34\dot\imath,\; \lambda_{2}=2.15 \dot\imath
         ,\;c=-0.2$. }
        \label{fig:figure3}
\end{figure}

For two soliton, take the matrices $\mathcal I^{(2)},\;\mathcal
\Theta_{n,\;1}^{m},\; \Theta_{n,\;2}^{m},\;\Lambda_{1}$ and
$\Lambda_{2}$ to be
\begin{eqnarray}\label{cccc5}
&& \mathcal{I}^{(2)}=\left(\begin{array}{cc}
1 & 0 \\
0 & 1 \\
0 & 0 \\
0 & 0 \\
\end{array}%
\right), \notag \\
&& \Theta_{n,\;1}^{m}=\left(
\begin{array}{cc}
 {\theta^{m}_{n,\;11}}^{(1)} &  {\theta^{m}_{n,\;12}}^{(2)} \\
 {\theta^{m}_{n,\;21}}^{(1)} &  {\theta^{m}_{n,\;22}}^{(2)}
\end{array}%
\right), \quad \Theta_{n,\;2}^{m}=\left(
\begin{array}{cc}
 {\theta^{m}_{n,\;11}}^{(3)} &  {\theta^{m}_{n,\;12}}^{(4)} \\
 {\theta^{m}_{n,\;21}}^{(3)} &  {\theta^{m}_{n,\;22}}^{(4)}
\end{array}%
\right), \notag \\
&&\Lambda_{1}=\left(\begin{array}{cc}
\lambda_{1} & 0 \\
0 & \bar{\lambda}_{1} \\
\end{array}%
\right), \quad \qquad \Lambda_{2}=\left(\begin{array}{cc}
                  \lambda_{2} & 0 \\
                  0 & \bar{\lambda}_{2} \\
                \end{array}%
              \right),
\end{eqnarray}
so that the matrices ${\mathcal G}_{n}^{m}$ and $\widetilde{\mathcal
G}_{n}^{m}$ become
\begin{eqnarray}\label{cccc6}
{\mathcal G}_{n}^{m} &=& \left( \begin{array}{cc} \Theta_{n,\;1}^{m} &  \Theta_{n,\;2}^{m} \\
 \Theta_{n,\;1}^{m}\Lambda_{1}^{-1} & \Theta_{n,\;2}^{m}\Lambda_{2}^{-1}
\end{array} \right)
=\left( \begin{array}{c:c}
\begin{array}{cc}
 {\theta^{m}_{n,\;11}}^{(1)} &  {\theta^{m}_{n,\;12}}^{(2)} \\
 {\theta^{m}_{n,\;21}}^{(1)} &  {\theta^{m}_{n,\;22}}^{(2)}
\end{array} & \begin{array}{cc}
  {\theta^{m}_{n,\;11}}^{(3)} &  {\theta^{m}_{n,\;12}}^{(4)} \\
 {\theta^{m}_{n,\;21}}^{(3)} &  {\theta^{m}_{n,\;22}}^{(4)}
              \end{array} \\  \hdashline
              \begin{array}{cc}
  \lambda_{1}^{-1}{\theta^{m}_{n,\;11}}^{(1)} &  \bar{\lambda}_{1}^{-1}{\theta^{m}_{n,\;12}}^{(2)} \\
 \lambda^{-1}{\theta^{m}_{n,\;21}}^{(1)} &  \bar{\lambda}_{1}^{-1}{\theta^{m}_{n,\;22}}^{(2)}
\end{array} &
\begin{array}{cc}
  \lambda_{2}^{-1}{\theta^{m}_{n,\;11}}^{(3)} &  \bar{\lambda}_{2}^{-1}{\theta^{m}_{n,\;12}}^{(4)} \\
 \lambda_{2}^{-1}{\theta^{m}_{n,\;21}}^{(3)} &  \bar{\lambda}_{2}^{-1}{\theta^{m}_{n,\;22}}^{(4)}
\end{array}
\end{array}%
\right), \notag \\
\widetilde{\mathcal{G}}_{n}^{m} &=& \left( \begin{array}{cc}
\Theta_{n,\;1}^{m}\Lambda_{1}^{-2} &
\Theta_{n,\;2}^{m}\Lambda_{2}^{-2}
\end{array} \right) =\left( \begin{array}{c:c}
\begin{array}{cc}
  \lambda_{1}^{-2}{\theta^{m}_{n,\;11}}^{(1)} &  \bar{\lambda}_{1}^{-2}{\theta^{m}_{n,\;12}}^{(2)} \\
 \lambda_{1}^{-2}{\theta^{m}_{n,\;21}}^{(1)} &  \bar{\lambda}_{1}^{-2}{\theta^{m}_{n,\;22}}^{(2)}
\end{array} &
\begin{array}{cc}
\lambda_{2}^{-2}{\theta^{m}_{n,\;11}}^{(3)} &  \bar{\lambda}_{2}^{-2}{\theta^{m}_{n,\;12}}^{(4)} \\
 \lambda_{2}^{-2}{\theta^{m}_{n,\;21}}^{(3)} &  \bar{\lambda}_{2}^{-2}{\theta^{m}_{n,\;22}}^{(4)}
\end{array}
\end{array}%
\right). \notag
\end{eqnarray}
The two-fold scalar solutions $u_{n}^{m}[2]$ and $v_{n}^{m}[2]$ are
given by
\begin{eqnarray}
u_{n}^{m}[2] &=& \dot\imath c - ({\mathcal Q^{m}_{n+1,\;11}}^{(2)} -
{\mathcal Q^{m}_{n,\;11}}^{(2)}), \label{twofoldsol1} \\
v_{n}^{m}[2] &=& - ({\mathcal Q^{m}_{n+1,\;12}}^{(2)} - {\mathcal
Q^{m}_{n,\;12}}^{(2)}). \label{twofoldsol2}
\end{eqnarray}
By using (\ref{b2}), one can compute the matrix elements ${\mathcal
Q^{m}_{n,\;11}}^{(2)},\;{\mathcal Q^{m}_{n,\;12}}^{(2)}$ as follow
\begin{eqnarray}\label{the11twofold}
{\mathcal Q^{m}_{n,\;11}}^{(2)} &=& \left\vert
\begin{array}{cc}
{\mathcal G}_{n}^{m} & {\mathcal I}^{(2)}_{1} \\
(\widetilde{\mathcal G}_{n}^{m})_{1} & \fbox{$ O $}
\end{array}%
\right \vert = \left\vert  \begin{array}{ccccc}
 {\theta^{m}_{n,\;11}}^{(1)} &  {\theta^{m}_{n,\;12}}^{(2)} & {\theta^{m}_{n,\;11}}^{(3)} &  {\theta^{m}_{n,\;12}}^{(4)} & 1 \\
 {\theta^{m}_{n,\;21}}^{(1)} &  {\theta^{m}_{n,\;22}}^{(2)} & {\theta^{m}_{n,\;21}}^{(3)} &  {\theta^{m}_{n,\;22}}^{(4)} & 0 \\
  \lambda_{1}^{-1}{\theta^{m}_{n,\;11}}^{(1)} & \bar{\lambda}_{1}^{-1}{\theta^{m}_{n,\;12}}^{(2)} & \lambda_{2}^{-1}
  {\theta^{m}_{n,\;11}}^{(3)} & \bar{\lambda}_{2}^{-1}{\theta^{m}_{n,\;12}}^{(4)} & 0 \\
\lambda_{1}^{-1}{\theta^{m}_{n,\;21}}^{(1)} &
\bar{\lambda}_{1}^{-1}{\theta^{m}_{n,\;22}}^{(2)} & \lambda_{2}^{-1}
  {\theta^{m}_{n,\;21}}^{(3)} & \bar{\lambda}_{2}^{-1}{\theta^{m}_{n,\;22}}^{(4)} & 0 \\
\lambda_{1}^{-2}{\theta^{m}_{n,\;11}}^{(1)} &
\bar{\lambda}_{1}^{-2}{\theta^{m}_{n,\;12}}^{(2)} & \lambda_{2}^{-2}
  {\theta^{m}_{n,\;11}}^{(3)} & \bar{\lambda}_{2}^{-2}{\theta^{m}_{n,\;12}}^{(4)}  & \fbox{$0$}
\end{array}%
 \right \vert, \notag \\ &=& - \frac{\det
\left(
\begin{array}{cccc}
 \lambda_{1}^{-2}{\theta^{m}_{n,\;11}}^{(1)} &
\bar{\lambda}_{1}^{-2}{\theta^{m}_{n,\;12}}^{(2)} & \lambda_{2}^{-2}
  {\theta^{m}_{n,\;11}}^{(3)} & \bar{\lambda}_{2}^{-2}{\theta^{m}_{n,\;12}}^{(4)}  \\
 {\theta^{m}_{n,\;21}}^{(1)} &  {\theta^{m}_{n,\;22}}^{(2)} & {\theta^{m}_{n,\;21}}^{(3)} &  {\theta^{m}_{n,\;22}}^{(4)}  \\
  \lambda_{1}^{-1}{\theta^{m}_{n,\;11}}^{(1)} & \bar{\lambda}_{1}^{-1}{\theta^{m}_{n,\;12}}^{(2)} & \lambda_{2}^{-1}
  {\theta^{m}_{n,\;11}}^{(3)} & \bar{\lambda}_{2}^{-1}{\theta^{m}_{n,\;12}}^{(4)}  \\
\lambda_{1}^{-1}{\theta^{m}_{n,\;21}}^{(1)} &
\bar{\lambda}_{1}^{-1}{\theta^{m}_{n,\;22}}^{(2)} & \lambda_{2}^{-1}
  {\theta^{m}_{n,\;21}}^{(3)} & \bar{\lambda}_{2}^{-1}{\theta^{m}_{n,\;22}}^{(4)}
\end{array}%
\right)}{\det \left(
\begin{array}{cccc}
 {\theta^{m}_{n,\;11}}^{(1)} &
{\theta^{m}_{n,\;12}}^{(2)} &
  {\theta^{m}_{n,\;11}}^{(3)} & {\theta^{m}_{n,\;12}}^{(4)}  \\
 {\theta^{m}_{n,\;21}}^{(1)} &  {\theta^{m}_{n,\;22}}^{(2)} & {\theta^{m}_{n,\;21}}^{(3)} &  {\theta^{m}_{n,\;22}}^{(4)}  \\
  \lambda_{1}^{-1}{\theta^{m}_{n,\;11}}^{(1)} & \bar{\lambda}_{1}^{-1}{\theta^{m}_{n,\;12}}^{(2)} & \lambda_{2}^{-1}
  {\theta^{m}_{n,\;11}}^{(3)} & \bar{\lambda}_{2}^{-1}{\theta^{m}_{n,\;12}}^{(4)}  \\
\lambda_{1}^{-1}{\theta^{m}_{n,\;21}}^{(1)} &
\bar{\lambda}_{1}^{-1}{\theta^{m}_{n,\;22}}^{(2)} & \lambda_{2}^{-1}
  {\theta^{m}_{n,\;21}}^{(3)} & \bar{\lambda}_{2}^{-1}{\theta^{m}_{n,\;22}}^{(4)}
\end{array}%
\right)}.
\end{eqnarray}
Similarly
\begin{eqnarray}\label{the12twofold}
{\mathcal Q^{m}_{n,\;11}}^{(2)} &=& - \frac{\det \left(
\begin{array}{cccc}
 {\theta^{m}_{n,\;11}}^{(1)} &
{\theta^{m}_{n,\;12}}^{(2)} &
  {\theta^{m}_{n,\;11}}^{(3)} & {\theta^{m}_{n,\;12}}^{(4)}  \\
 \lambda_{1}^{-2}{\theta^{m}_{n,\;11}}^{(1)} &
\bar{\lambda}_{1}^{-2}{\theta^{m}_{n,\;12}}^{(2)} & \lambda_{2}^{-2}
  {\theta^{m}_{n,\;11}}^{(3)} & \bar{\lambda}_{2}^{-2}{\theta^{m}_{n,\;12}}^{(4)}   \\
  \lambda_{1}^{-1}{\theta^{m}_{n,\;11}}^{(1)} & \bar{\lambda}_{1}^{-1}{\theta^{m}_{n,\;12}}^{(2)} & \lambda_{2}^{-1}
  {\theta^{m}_{n,\;11}}^{(3)} & \bar{\lambda}_{2}^{-1}{\theta^{m}_{n,\;12}}^{(4)}  \\
\lambda_{1}^{-1}{\theta^{m}_{n,\;21}}^{(1)} &
\bar{\lambda}_{1}^{-1}{\theta^{m}_{n,\;22}}^{(2)} & \lambda_{2}^{-1}
  {\theta^{m}_{n,\;21}}^{(3)} & \bar{\lambda}_{2}^{-1}{\theta^{m}_{n,\;22}}^{(4)}
\end{array}%
\right)}{\det \left(
\begin{array}{cccc}
 {\theta^{m}_{n,\;11}}^{(1)} &
{\theta^{m}_{n,\;12}}^{(2)} &
  {\theta^{m}_{n,\;11}}^{(3)} & {\theta^{m}_{n,\;12}}^{(4)}  \\
 {\theta^{m}_{n,\;21}}^{(1)} &  {\theta^{m}_{n,\;22}}^{(2)} & {\theta^{m}_{n,\;21}}^{(3)} &  {\theta^{m}_{n,\;22}}^{(4)}  \\
  \lambda_{1}^{-1}{\theta^{m}_{n,\;11}}^{(1)} & \bar{\lambda}_{1}^{-1}{\theta^{m}_{n,\;12}}^{(2)} & \lambda_{2}^{-1}
  {\theta^{m}_{n,\;11}}^{(3)} & \bar{\lambda}_{2}^{-1}{\theta^{m}_{n,\;12}}^{(4)}  \\
\lambda_{1}^{-1}{\theta^{m}_{n,\;21}}^{(1)} &
\bar{\lambda}_{1}^{-1}{\theta^{m}_{n,\;22}}^{(2)} & \lambda_{2}^{-1}
  {\theta^{m}_{n,\;21}}^{(3)} & \bar{\lambda}_{2}^{-1}{\theta^{m}_{n,\;22}}^{(4)}
\end{array}%
\right)}.
\end{eqnarray}
The graphical representation of discrete soliton solution
(\ref{twofoldsol1}) of the time-discrete GLHM model has been
depicted in Figure \ref{fig:figure2}-\ref{fig:figure3}
To obtain three-soliton solution, we take three particular matrix
solutions $\Theta_{n,\;k}^{m}$ corresponding to eigenvalue matrices
$\Lambda_{k},\; (k = 1,\; 2,\; 3)$. Figures
\ref{fig:figure2}-\ref{fig:figure3} describes the interactions of
two discrete solitons with their own lumps of energies moving with
different velocities. These independent solitons propagate in space
and keeps their profiles unchanged before and after collision. After
collision, they separate and travel through each other independently
and retains their amplitudes and velocities invariant. It is to be
noted that, the shape of these solitons are characterized under
given parametric conditions. When these conditions changes, the
structure of the soliton can also be changed. Similarly, by an
application of DT K-times on a seed solution, one can obtain
K-soliton (or multisoliton) solutions of time-discrete GLHM model.
\section{Concluding remarks}
In this paper, we have studied Darboux transformation and soliton
solutions of time-discrete GLHM model. We have defined Darboux
transformation on the solution to the Lax pair and the solutions of
time-discrete GLHM model. The solutions are expressed in terms of
quasideterminants. Finally, soliton solutions of the time-discrete
GLHM model are calculated for the general and simple cases. We have
computed expressions of one-soliton solution by expanding
quasideterminants. Further, we remarks here that
\begin{enumerate}
\item[$\bullet$] The time-discrete GLHM model
studied in the present paper can also be served as a numerical
scheme for the numerical simulation of the continuous time GLHM
model.
\item[$\bullet$] The work can be further extended by studying
Hirota bilinearization of the time-discrete GLHM model.
\end{enumerate}

\end{document}